# Quantum Computing based Hybrid Solution Strategies for Large-scale Discrete-Continuous Optimization Problems

Akshay Ajagekar[1], Travis Humble[2], Fengqi You[*1]

[1] Robert Frederick Smith School of Chemical and Biomolecular Engineering,

Cornell University, Ithaca, New York 14853, USA

[2] Quantum Computing Institute, Oak Ridge National Laboratory,

Oak Ridge, Tennessee 37831, USA

## Abstract

Quantum computing (QC) has gained popularity due to its unique capabilities that are quite different from that of classical computers in terms of speed and methods of operations. This paper proposes hybrid models and methods that effectively leverage the complementary strengths of deterministic algorithms and QC techniques to overcome combinatorial complexity for solving large-scale mixed-integer programming problems. Four applications, namely the molecular conformation problem, job-shop scheduling problem, manufacturing cell formation problem, and the vehicle routing problem, are specifically addressed. Large-scale instances of these application problems across multiple scales ranging from molecular design to logistics optimization are computationally challenging for deterministic optimization algorithms on classical computers. To address the computational challenges, hybrid QC-based algorithms are proposed and extensive computational experimental results are presented to demonstrate their applicability and efficiency. The proposed QC-based solution strategies enjoy high computational efficiency in terms of solution quality and computation time, by utilizing the unique features of both classical and quantum computers.

---

[*]Corresponding author. Phone: (607) 255-1162; Fax: (607) 255-9166; E-mail: fengqi.you@cornell.edu

## 1. Introduction

Computational optimization is a ubiquitous paradigm with a wide range of applications in science and engineering, and it has received tremendous attention from both academia and industry [1]. For chemical engineering applications, especially in process systems engineering, optimization is an integral part of synthesis, design, operations, and control [2]. Large-scale optimization problems with complex economic and performance interactions could be computationally intractable to solve through off-the-shelf methods and might require specialized solution algorithms. Therefore, it is important to explore advanced computational paradigms and strategies that can address complex, large-scale optimization problems.

Primary components of the optimization process are the computational modeling and search algorithms [1]. Growing demand for optimization techniques that can obtain an optimal solution, requires the search algorithms to be efficient and computationally tractable. Over the past years, deterministic optimization techniques, especially nonlinear programming and mixed-integer nonlinear programming (MINLP) algorithms, have received growing attention from academia and industry [3, 4]. Despite the algorithmic and applications-oriented advances in global optimization [5], large-scale nonconvex MINLP problems could still be computationally very expensive to solve with the state-of-the-art deterministic global optimization algorithms. Specifically, deterministic algorithms for solving large-scale MINLP problems need to deal with the growing size of sub-problems and/or exponential growth of branch-and-bound tree, leading to enumeration of many more alternatives in the feasible space. Heuristic techniques such as simulated annealing [6], genetic algorithms [7], tabu search [8] and others have also grown popular owing to their easy implementation and little prior knowledge requirement of the optimization problem. However, such techniques are most suitable for unconstrained optimization problems and show no rigorous convergence properties. The ever-increasing complexity of combinatorial optimization problems accompanied by a quickly growing search space, results in the optimization process being more time-consuming. This computational challenge could be handled by advancements in computers with high processing speeds, but saturation limits of Moore's law render the possibility of rising processing speeds unlikely in the coming years [9]. There arises a need for novel solution approaches capable of overcoming limitations of the current optimization paradigms carried out on state-of-the-art classical computers.

Quantum computing (QC) is the next frontier in computation and has attracted a lot of attention from the scientific community in recent years. QC provides a novel approach to help solve some of the most complex optimization problems while offering an essential speed advantage over classical methods [10]. This is evident from QC techniques like Shor's algorithm for integer factorization [11], Grover's search algorithm for unstructured databases [12], quantum algorithm for linear system of equations [13], and many more [14]. Quantum adiabatic algorithms too are efficient optimization strategies that quickly search over the solution space [15]. Quantum computers perform computation by inducing quantum speedups whose scaling far exceeds the capability of the most powerful classical computers. QC's major applications can be perceived in areas of optimization, machine learning, cryptography, and quantum chemistry [16]. Despite the contrasting views on QC's viability and performance, there is no doubt that QC holds great promise to open up a new era of computing.

QC-based solution approaches are in their earliest stages of development compared to their much more matured classical counterparts. Current quantum machines have very limited functionality in the context of optimization, such that QC hardware and algorithms are inadequate for large-scale optimization problems. Although it has been shown that some optimization problems relevant to energy systems can be solved using quantum computers, their performance deteriorates with increasing size and complexity [17]. A number of technological limitations face commercially available quantum computers, such as relatively small number of qubits with limited connectivity, and lack of quantum memory. Therefore, harnessing the complementary strengths of classical and quantum computers to solve complex large-scale optimization problems has become the main strategy for near-term and mid-term solution [18, 19].

There are several research challenges towards developing hybrid QC-based solution strategies for large-scale mixed-integer optimization problems. The first challenge is to develop a hybrid algorithmic framework that leverages both QC and classical computers, by integrating exact solution techniques with QC-based solution techniques. A further challenge lies in developing subproblems that can be solved using QC techniques, with an essential advantage of computation speed over classical solution methods. The formulation of these subproblems directly depends on the properties and structure of the original problem. Formulating appropriate subproblems that can be solved on a quantum computer poses another research challenge. The final challenge is to keep the resulting subproblems small enough such that they can be run on current quantum systems. For example, QC-based solution algorithms for operational planning problems can only solve small-scale instances and do not produce results that are comparable with those obtained from the state-of-the-art classical computing approaches [20, 21]. Developing an algorithmic framework that can benefit from high-quality solutions obtained through QC techniques, is crucial to overcome such computational challenge.

The objective of this paper is to develop hybrid QC-based models and methods that exploit the complementary strengths of QC and exact solution techniques to overcome the combinatorial complexity when solving large-scale discrete-continuous optimization problems. Discrete-continuous optimization problems are harder to solve than continuous optimization problems due to the combinatorial explosion that occurs in all but smaller problems. The high combinatorial complexity stemming from this explosion in large-scale discrete-continuous optimization problems can be tackled by QC-based solution techniques. In addition, since quantum computers can only handle discrete binary variables, large-scale optimization problems that involve both discrete decision variables and continuous variables are most favorable to be solved through hybrid QC-based solution strategies. The applicability of these QC-based algorithms is demonstrated by large-scale applications across scales that are relevant to molecular design, process scheduling, manufacturing systems operations, and vehicle routing. Each application addressed in this paper belongs to a specific class of optimization problems. They include binary quadratic programming (BQP), mixed-integer linear programming (MILP), mixed-integer quadratic programming (MIQP), and integer quadratic fractional programming (IQFP) problems. The proposed hybrid QC-based solution methods effectively tackle the computational challenges stemming from the structures of the corresponding application problems. These computational challenges can arise from the large number of discrete variables, constraints and nonlinearities within an optimization problem. In order to demonstrate the computational efficiency of the proposed hybrid QC-based methods, large-scale instances of each application problem are solved using the proposed hybrid solution techniques. The obtained computational results are compared against the results obtained with general-purpose state-of-the-art optimization solvers that are implemented on classical computers.

The novel contributions of this paper are summarized as follows:

- A QC-based method to solve molecular conformation problems using the hybrid QC partitioning algorithm;

- A novel hybrid QC-MILP decomposition method that obtains global optimal solutions for large-scale job-shop scheduling problems;

- A hybrid QC-MIQP stepwise decomposition method developed specifically for solving the manufacturing cell formation problem;

- A novel hybrid QC-IQFP parametric method that efficiently solves the vehicle routing problem that is formulated as an IQFP problem.

The remainder of this paper is organized as follows. Section 2 introduces some of the basic definitions and relevant properties of QC for computational optimization. A brief overview of the applications chosen for this study is presented in Section 3. It is followed by the application problems that are solved using the respective hybrid QC-based techniques and by off-the-shelf deterministic optimization solvers as well. Conclusions are drawn in Section 8.

## 2. Background on Quantum Computing for Optimization

Unlike the classical computers, quantum computers follow the logic of quantum mechanics. The fundamental unit of quantum computer is called a *quantum bit* or a *qubit*. The quantum state of qubits can be in superposition of their basis states, setting them apart from classical bits which can be in either of the two discrete states. Although infinite quantum states are possible for qubits, they collapse into one of their basis states after measurement [10]. Another elegant property of qubits is their ability to form entangled states with each other, allowing them to form co-relations between individually random behaviors. A two-qubit system is shown in Figure 1a. Quantum computers exploit the qubit properties of superposition and entanglement to perform computations.

Two QC architectures with fundamental differences in their operations are commercially available. They are the gate-model quantum computer and the annealing-based quantum computer [22, 23]. The gate model as seen in Figure 1b uses quantum gates to manipulate qubit states and perform calculations. Quantum gate operations are sequentially applied to qubit states and evolve them towards a desired solution of the problem [24]. On the other hand, the annealing-based model intrinsically realizes the quantum annealing algorithm for its operation. Quantum annealing is a quantum analogue of classical simulated annealing that permits quantum tunneling as shown in Figure 1c to aid in exploring low-cost solutions and ultimately yield a global minimum [25]. Quantum annealing also exhibits convergence to the optimal or ground state with larger probability than simulated annealing [25]. Although an extended form of quantum annealing is theoretically equivalent to the gate model [26], their modes of operation are quite different.

Commercial QC architectures are in the rudimentary stages of development and exhibit limitations in terms of ease of computation, performance, and even algorithmic limitations. Factors like poor error correction, qubits susceptible to de-coherence, and limited quantum control contribute towards the aforementioned hindrances. From an optimization perspective, annealing-based quantum computers are much closer to discrete optimization problems than the gate-model quantum computers [27]. This is due to the fact that annealing-based devices are built explicitly for optimization facilitated by quantum annealing, while a gate-model quantum computer follows a universal computation approach. Additionally, due to fewer number of qubits with poor error correction, the current gate-model quantum computers exhibit poor performance in terms of computation time required to find an optimal solution. Weighing the strengths and drawbacks of annealing-based and gate-model QC devices, the annealing-based quantum computers are much more reliable when solving optimization problems. Therefore, our work focuses on quantum annealing-based optimization techniques for several complex combinatorial optimization problems of practical relevance. It is also interesting to note that the size of problems that can be solved on quantum annealing machines have continued to grow as the hardware capacity has increased, since the physical realization of such quantum systems back in 2011 [28].

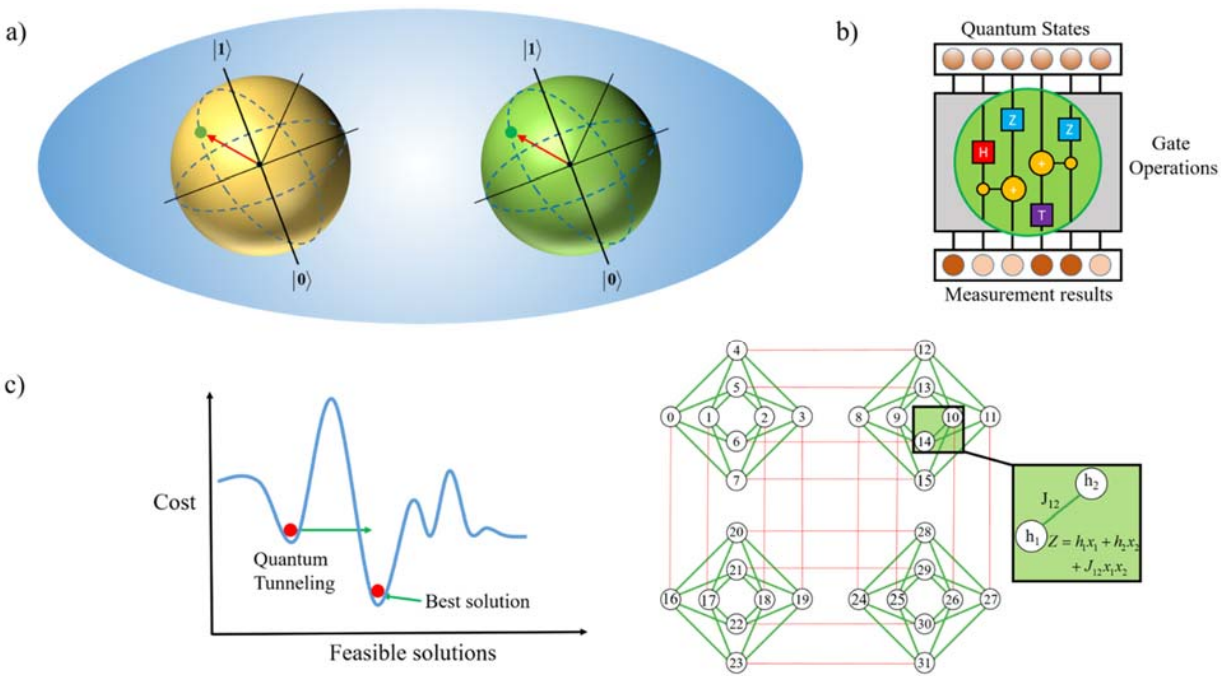

*Figure 1. a) Entangled qubits in superposition states, b) gate-model quantum computer operation, and c) quantum annealing-based computer operation*

Quantum annealing is an elegant approach that helps escape local minima and overcomes barriers by tunneling through them rather than stochastically overcoming them. In quantum annealing, the system is initialized in the lowest-energy eigenstate of the initial Hamiltonian. A Hamiltonian is a mathematical description of the physical system in terms of its energies and corresponds to the objective function of an optimization problem [29]. The annealing process then proceeds by evolving the quantum state towards a user-defined problem Hamiltonian for the system. The influence of the initial Hamiltonian is also reduced adiabatically to yield an eigenstate of the problem Hamiltonian when the annealing schedule ends [25]. The amplitude of initial Hamiltonian causes quantum tunneling between various classical states or the eigenstates of the problem Hamiltonian. By decreasing this amplitude from a very large value to zero, the system is driven into the optimal state that is the ground state of the problem Hamiltonian. Compared to its classical counterpart, quantum annealing gives a larger probability to lead to the ground state under the same conditions on the annealing schedule and interactions [25].

A family of commercially available QC devices from D-Wave Systems are designed to implement quantum annealing. In order to solve problems on the D-Wave system, they need to be formulated as an Ising model or, equivalently, a quadratic unconstrained binary optimization (QUBO) model. These models can be further represented by a graph comprising a collection of nodes and edges between them while the corresponding quantum processing unit is expressed as a lattice of qubits interconnected in a design known as a *Chimera* graph. Figure 1c represents a subgraph of the *Chimera* lattice pattern that is typical of the D-Wave quantum annealing systems and its operation. The nodes and edges of the objective function graph are mapped to the qubits and couplers of the *Chimera* lattice. Since the *Chimera* lattice is not fully connected, mapping of variables to physical qubits uses the process of minor-embedding and is crucial to problem solving [30-33]. While the problem of finding an optimal graph-minor is itself NP-hard, an efficient embedding for many graphs can be found systematically with heuristic techniques [34]. Mapping the objective function onto the physical quantum processing unit is followed by the realization of quantum annealing process [35] which searches for low-energy solutions of the corresponding problem Hamiltonian [36]. The probability of recovering the global optimal solution is highly dependent on the embedding and annealing schedule [37, 38].

## 3. Overview of Quantum Optimization Application Domains

In this section, we provide a brief overview of the application problems addressed in this work, as well as their model formulations and the proposed QC-based solution algorithms (see Table 1). Four applications are considered in this paper, namely the molecular conformation problem [39], the job-shop scheduling problem [40], the manufacturing cell formation problem and the vehicle routing problem [41]. These applications cover applications across many different scales, from molecular design to process operations and to supply chain and logistics optimization. The order of these application problems is also arranged such that their complexity increases successively.

The molecular conformation problem is a building block of molecular design, having major implications in the field of drug design and product design [42]. Section 4 describes the molecular conformation problem. It is followed by the job-shop scheduling problem in Section 5. The job-shop scheduling problem is a notoriously difficult problem in combinatorial optimization and forms the basis of several practical production scheduling problems [43]. Cell formation is an important component of cellular manufacturing, and has been gaining popularity in manufacturing industries as well as engineering management [44]. We consider one of the several formulations of the manufacturing cell formation problem in Section 6. One of the well-studied problems in logistics is the vehicle routing problem which has a large number of real-world applications. In Section 7, we consider a nonlinear formulation of the vehicle routing problem. It should also be noted that the scope of the proposed hybrid algorithms presented in Table 1 is not limited to these examples, but can be extended to other real-world problems of practical relevance as well.

All the computational experiments are carried out on a Dell Optiplex system with Intel® Core™ i7-6300 3.40 GHz CPU and 32 GB RAM. Only one core is used for computation. The BQP, MILP, and MIQP problems are solved using Gurobi 8. The IQFP problems are solved using MINLP solvers Bonmin 15 and Baron 18. The same Optiplex computer is also used to perform basic programming functions and as a classical backend in conjunction with a quantum processor, for the hybrid QC-based solution methods. The D-Wave 2000Q quantum processor with 2,048 qubits and 5,600 couplers is used for all computational experiments involving hybrid QC-based methods. The quantum processor is set to use 1,000 reads and an anneal time of $20\mu s$.

*Table 1. Outline of applications presented in this work along with their distinguishing characteristics and solution methods*

| Application | Areas of applications | Objective | Formulation | Hybrid QC solution algorithm | General purpose solver |
|---|---|---|---|---|---|
| Molecular Conformation | Biotechnology, chemistry, drug design, and protein folding | Minimize potential energy | Binary Quadratic Program (BQP) | Existing hybrid QC partitioning approach | Gurobi 8 |
| Job-shop Scheduling | Production planning and scheduling | Minimize processing costs | Mixed-integer Linear Program (MILP) | Proposed hybrid QC-MILP decomposition method | Gurobi 8 |
| Manufacturing Cell Formation | Cellular manufacturing, group technology | Minimize manufacturing costs | Mixed-integer Quadratic Program (MIQP) | Proposed hybrid QC-MIQP stepwise decomposition method | Gurobi 8 |
| Vehicle Routing | Transportation systems, production planning and logistics, waste management | Minimize ratio of travel costs to resources spent | Integer Quadratic Fractional Program (IQFP) | Proposed hybrid QC-IQFP parametric method | Bonmin 15, Baron 18 |

## 4. QC for Molecular Conformations in Molecular/Product Design

Any spatial arrangement of the atoms in a molecule that result from rotations about their single bonds are termed as molecular conformations. Molecules in nature may change their conformation as a stimulus to change in the surrounding environmental conditions. These environmental conditions drive changes in potential energy function for a cluster of atoms. Minimization of total potential energy associated with configuration of atoms in a molecule is known as the molecular conformation problem [45]. An important area of research in computational biochemistry and biotechnology is the design of molecules for specific applications, such as determining the protein folded state from a known primary sequence of amino acids. Such applications require solving the molecular conformation problem to global optimality [46]. Molecular conformation problem helps predict the native structure of sequence of molecules, and it can be considered as a simplified form of predicting native structures of residues in case of protein folding [39, 47-49]. Heuristic solution techniques for the molecular conformation problem are computationally expensive, depend heavily on carefully chosen parameters, and do not guarantee a global optimum [50]. Exact techniques to solve the molecular conformation problem have also been proposed, but they could perform poorly as the size of the molecule increases [45, 51].

To eliminate complex nonlinearities, the molecular conformation problem can be modeled by a discrete approximation on a 3-dimensional lattice. Solution to this simplified discretization of the molecular conformation problem might not provide a global optimum for the original continuous and highly nonlinear molecular design problem. However, this solution can serve as a starting point for the global optimization of the continuous molecular conformation problem [47]. This approach has been successful in finding the minimum energy conformations for very large molecules [52, 53]. QC-based solution techniques can help solve the discretized molecular conformation problems efficiently by offering a quantum advantage in terms of computation speed. Quantum annealing searches the molecular conformation space with the promise of returning a configuration that is near the global minimum.

## 4.1. Model Formulation

A molecule comprising of $B$ atoms modeled as single *spheres* or *beads* is placed inside a 3-dimensional cubic lattice with $N$ sites such that $N \geq B$. In a string of beads model, the molecule consists of $B$ beads, $a_1, a_2,..a_B$, where $a_i$ denotes the $i$th bead in the primary sequence. Between every pair of consecutive beads $a_i$ and $a_{i+1}$, there exists a bond of length $lb_i$. The binary assignment variable $x_{ij}$ represents the assignment of bead $i$ at lattice site $s_j$. Pairwise potential between beads $a_i$ and $a_k$ placed at sites $s_j$ and $s_l$ is modeled as Leonard-Jones (LJ) potential given by $U_{ijkl}^{LJ}$ in Eq. (1), where $\varepsilon_{ik}$ and $\sigma_{ik}$ are LJ parameters representing the depth of potential well and distance at which inter-particle potential is zero, respectively. These parameter values are dependent on the nature of beads.

$$U_{ijkl}^{LJ} = 4\varepsilon_{ik}\left(\left(\frac{\sigma_{ik}}{r_{jl}}\right)^{12} - \left(\frac{\sigma_{ik}}{r_{jl}}\right)^{6}\right) \tag{1}$$

A bond stretching potential $U_{ijkl}^{bond}$ is introduced between each consecutive pair of beads $a_i$ and $a_k$ given in Eq. (2). Distance between lattice sites $s_j$ and $s_l$ is given by $r_{jl} = \left\| s_j - s_l \right\|_2$. The penalty parameter $\beta$ enforces that the distances between consecutive beads remain within an allowable distance of the required bond lengths. $U_{ijkl}$ represents the total potential energy contribution to the free energy of the system due to placement of beads $a_i$ and $a_k$ at sites $s_j$ and $s_l$. Moreover, the terms $U_{ijil}$ and $U_{ijkj}$ are set to a very high value to ensure that no two beads are placed at the same location, and no bead is assigned to two locations. Bond bending potentials and torsional potentials are ignored in this case due to their trivial contribution to the free energy of the system and for the sake of simplicity. The molecular conformation problem is to determine the locations of beads within the cubic lattice, set of bond lengths, and bond angles.

$$U_{ijkl}^{bond} = \beta \left( r_{jl} - lb_i \right)^2 \tag{2}$$

$$U_{ijkl} = U_{ijkl}^{LJ} + U_{ijkl}^{bond} \tag{3}$$

The discretized molecular conformation problem can be formulated as a quadratic assignment problem [47]. The quadratic term $U_{ijkl} x_{ij} x_{kl}$ represents the direct contribution to total free energy when the bead $a_i$ and $a_k$ are assigned to sites $s_j$ and $s_l$, respectively. The objective function in Eq. (4) is the total potential energy of the system to be minimized. Constraints in (5) are assignment constraints to ensure that each bead occupies exactly one lattice site. Constraint (6) makes sure that at most one bead occupies each lattice site $s_j$.

$$\min \sum_{i}^{B} \sum_{j}^{N} \sum_{k}^{B} \sum_{l}^{N} U_{ijkl} x_{ij} x_{kl} \tag{4}$$

$$s.t. \quad \sum_{j=1}^{N} x_{ij} = 1, \quad \forall i = 1, 2, \dots B \tag{5}$$

$$\sum_{i=1}^{B} x_{ij} \leq 1, \quad \forall j = 1, 2, \dots N \tag{6}$$

$$x_{ij} \in \{0, 1\}, \quad \forall i = 1, \dots, B, \, \forall j = 1, \dots, N \tag{7}$$

## 4.2. Hybrid QC Partitioning Algorithm

As stated earlier in Section 2, only QUBO problems can be solved directly on the quantum annealing-based machine. Large-scale QUBO problem cannot be directly fit on modest-sized *Chimera* lattice, and a specialized hybrid solution approach based on partitioning is required. The hybrid QC partitioning algorithm is based on a two-level approach. The full QUBO problem is the primary level, and the secondary level is a sub-QUBO problem sized to fit in the available quantum processing unit [54]. This algorithm exploits the complementary strengths of the quantum solver and classical tabu search. It can be viewed as a large-neighborhood local search with tabu improvements after each iteration.

Each iteration of the hybrid QC partitioning algorithm comprises of multiple calls to the quantum computer to globally minimize each sub-QUBO and a tabu search call for local minimization. The key idea behind the hybrid algorithm revolves around determining an order of variables in the QUBO problem based on their impact on the objective function value. First, the QUBO problem is split into sub-QUBOs that can be fit on the quantum processor, and then solved to optimality following the minor-embedding process. The solution vector is updated with the appropriate variable values from the sub-QUBO solution vectors, such that the updated solution vector jumps out of a local minimum. The new solution is passed on to tabu search, in order to obtain a new local minimum. This process is repeated until no better solution is found. The open-source software tool *qbsolv* implements this hybrid QC partitioning algorithm to solve QUBO problems [54].

$$\min H = \sum_{i}^{B} \sum_{j}^{N} \sum_{k}^{B} \sum_{l}^{N} U_{ijkl} x_{ij} x_{kl} + A \sum_{i=1}^{B} \left( \sum_{j=1}^{N} x_{ij} - 1 \right)^2 + A \sum_{j=1}^{N} \left( \sum_{i=1}^{B} x_{ij} \left( \sum_{k=1}^{B} x_{kj} - 1 \right) \right) \tag{8}$$

The molecular conformation problem can be formulated as a QUBO problem by modeling the assignment constraints as weighted penalty functions. Eq. (8) represents the QUBO formulation of the Hamiltonian to be minimized for the molecular conformation problem. The weight parameter $A$ is chosen such that $A \gg U_{ijkl}$ to enforce constraint satisfaction. It should also be noted that the size of this dense QUBO problem increases quadratically with the number of binary variables. The hybrid QC partitioning algorithm solves the discretized molecular conformation problem formulated as a QUBO problem by partitioning it into smaller sub-QUBO problems that can be efficiently solved on the quantum computer.

## 4.3. Computational Results

In order to illustrate the viability of the discretized formulation of the molecular conformation problem, we perform computational experiments for the alkane molecule named butane. This example is borrowed from literature along with its LJ parameters [47]. The butane molecule comprises of four carbon atoms and can exist in one of the four conformation states. The anti conformation shown in Figure 2a is the most stable and well-known conformer of butane with the lowest potential energy. The molecular conformation problem for butane is initialized by creating a 4×4×4 cubic lattice with a unit cell of length 1.4 A. The resulting formulation consists of 4 beads to be placed among 64 lattice sites, implying 256 binary variables in the problem along with 68 constraints. The potential energy contributions are calculated for each pair of bead and lattice site with the LJ parameter values as $\varepsilon = 0.06 \ kcal$ and $\sigma = 3.6 \ A$ [47]. The approximate bond length between the carbon atoms is known to be 1.5 A and can be used to calculate the bond penalty potentials in Eq. (2).

Solving the discretized molecular conformation problem for butane yields a gauche configuration as shown in Figure 2b. For butane, although the gauche conformer is less stable than the anti conformation, it is more stable than the other eclipsed configurations. Potential energy corresponding to the anti conformation is the global optimum of the continuous nonlinear molecular conformation problem, but the gauche conformation is a global optimum for the discretized problem. This disparity is due to the fact that the discretized molecular conformation problem makes several assumptions to reduce model complexities. However, it does not imply that the discretized molecular conformation model cannot be used in case of molecules found in nature. In fact, solutions obtained through the discretized molecular conformation model serve as a starting point for the global minimization of the more complex nonlinear molecular conformation problem over a continuous domain. This model can provide approximate solutions to the molecular conformation problems associated with most large molecules under study.

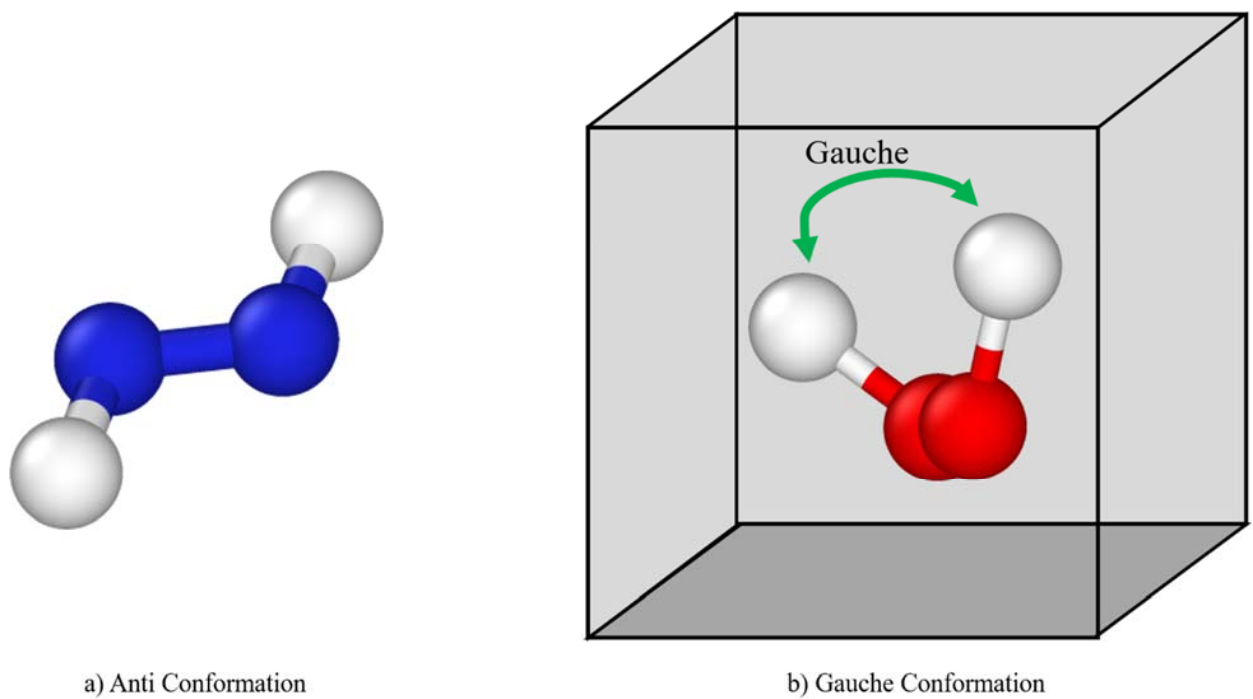

a) Anti Conformation                    b) Gauche Conformation

*Figure 2. a) Most stable anti conformation of butane, and b) obtained gauche conformation of butane*

In this section, we also consider molecular conformation instances with varying molecule sizes and lattice sizes. The instances are chosen such that each molecule is to be placed within different sized cubic lattices. Theoretically, the minimum energy for a molecular conformation in a larger lattice is expected to be less than or equal to that of the energy of the same conformation in a smaller lattice. This follows only if the cubic lattice is large enough to contain the entire molecule. For each of these instances, the LJ parameters are set to unity and potential energy contribution between each pair of atom and lattice site is calculated beforehand. The problem sizes of instances for $B$ atom beads and $N$ lattice sites are given in Table 1. This means that, for the largest problem with 12 atoms and 1,000 lattice sites, the BQP problem includes 12,000 binary variables and 1,012 constraints, and the reformulated QUBO problem includes 12,000 binary variables with no constraints. All the integer programming formulations of molecular conformation problem are modeled with Pyomo [55] and solved with MILP solver Gurobi. The corresponding QUBO formulations are solved using the aforementioned hybrid QC partitioning technique through the Python based tool *qbsolv.* A time limit of 24 hours in Gurobi is set for each molecular conformation instance to enforce appropriate comparison with the hybrid QC partitioning approach. Classical computing and QC facilities reported in Section 3 are used to carry out all the computational experiments.

Table 2 shows the objective function values representing the potential energy of the molecular configuration for solving problems directly using Gurobi and the hybrid QC partitioning algorithm. The detailed computational times of the computational experiments using different methods to solve molecular conformation instances are also presented in Table 2. The instances are chosen such that they range across molecules containing 3 atoms to 12 atoms. As seen in the table, small instances that correspond to small molecules placed in a small cubic lattice are solved to optimality by the deterministic MILP solver Gurobi. Least energy solutions are also obtained by the hybrid QC partitioning algorithm for the same molecular conformation instances but require longer runtimes. As evident from the computational results, deterministic solver Gurobi solves small instances within much shorter computation times than the hybrid QC partitioning algorithm. The smallest problem considered here with 81 binary variables is too large to be directly embedded on the quantum computer and therefore requires partitioning into smaller subproblems by the hybrid QC partitioning algorithm. Although these subproblems can be solved much faster on the quantum computer, the higher computation time account for solving the large number of subproblems formed. However, as the sizes of instances increase, Gurobi solver is unable to obtain optimal solutions with zero optimality gap. A minimum energy conformation for molecules containing more than 8 atoms when placed in lattice size larger than 6 units, cannot be found using the branch-and-cut algorithm implemented in Gurobi solver after running for 24 hours. The largest instance with 12 atoms cannot be solved to global optimality due to the physical memory limitations of the classical computer and hence yields a suboptimal solution. The hybrid QC partitioning approach on the other hand, finds "good quality" near-optimal solutions for these relatively large instances within reasonable computational time. Since global optimal solutions are not available for the larger molecular conformation problems, only conjecture can be made on the least energy solutions for each of these instances. For a molecule of a specific size, increasing the corresponding lattice size yields a conformation with potential energy slightly higher than that of the smaller lattice, through the hybrid QC partitioning based technique. This indicates that the optimal solution obtained through a hybrid QC approach lies within a local neighborhood of its global minimum. The computational results show that the QUBO problems solved through the hybrid QC partitioning algorithm are quite effective in terms of both computation time and solution quality.

The performance of the hybrid QC partitioning based approach is competitive in numerical results and superior in computation time, compared to the classical deterministic solver Gurobi for the molecular conformation problem. Being a heuristic technique, the hybrid QC partitioning approach does not always guarantee a lowest-energy solution. However, this hybrid approach can provide a "good quality" upper bound and a starting point for the continuous molecular conformation problem, and this can prove beneficial when good enough solutions are expected within short computation times.

*Table 2. Computational results of the molecular conformation problems*

| Molecular size | | Binary variables | Gurobi 8 | | Hybrid QC partitioning algorithm | |
|---|---|---|---|---|---|---|
| Atoms | Lattice sites | | Time (s) | Min obj. | Time (s) | Min obj. |
| 3 | 27 | 81 | 0.14 | -0.17 | 173 | -0.17 |
| 4 | 27 | 108 | 0.10 | -0.33 | 574 | -0.33 |
| 5 | 125 | 625 | 646 | -0.42 | 1,010 | -0.42 |
| 5 | 216 | 1,080 | 3,820 | -0.42 | 1,797 | -0.42 |
| 8 | 216 | 1,728 | 86,400* | -0.83 | 2,589 | -0.76 |
| 8 | 512 | 4,096 | 86,400* | -0.83 | 4,762 | -0.68 |
| 10 | 216 | 2,160 | 86,400* | -1.08 | 2,283 | -0.85 |
| 10 | 512 | 5,120 | 86,400* | -1.08 | 9,009 | -0.79 |
| 12 | 512 | 6,144 | 86,400* | -1.33 | 10,384 | -0.94 |
| 12 | 1,000 | 12,000 | 86,400* | -1.33 | 27,694 | -0.93 |

* Timeout of 24 hours (86,400 CPUs) reached and the best solution found is reported.

## 5. QC for Job-shop Scheduling

Job-shop scheduling problems belong to the class of most intractable NP-hard combinatorial optimization problems, and pose a significant computational challenge due to its large and complex search space [43]. The goal of the job-shop scheduling problem is to schedule a set of jobs on a set of machines subject to operational, scheduling, and logic constraints, in order to minimize the total processing cost. Advances in its solution algorithms have been focused on the development of exact methods like branch-and-bound that make use of bounds based on Lagrangean relaxation, bounds based on valid inequalities, and cutting planes [56]. Several heuristic solution techniques have also been proposed to solve complex job-shop scheduling problems, but they are unable to find high-quality solutions if large-size and complex search spaces are involved [56, 57]. Decomposition-based algorithms and hybrid algorithms that combine exact solution methods with constraint programming reduce the combinatorial complexity of job-shop scheduling problems, and have proven effective for several real-world industrial-scale problems [58, 59].

## 5.1. Model Formulation

In this section, we consider the MILP model of the job-shop scheduling problem with due dates and sequence-independent processing times [59]. This single-stage parallel scheduling problem considers a set of jobs $I$ using a set of machines $M$. Processing job $i \in I$ on machine $m \in M$ requires $P_{im}$ amount of time and costs $C_{im}$. Job $i \in I$ can only begin after the release date, and must be completed before its due date represented by $R_i$ and $D_i$, respectively. The processing costs, processing times, and release and due dates for each job-machine pair are known beforehand, and are independent of the sequence.

The decision variables in this MILP model are $ts_i$, $x_{im}$ and $y_{ij}$, representing the start time of jobs, assignments, and sequence of jobs on each machine, respectively. Binary variables $x_{im}$ are assignment variables that indicate whether job $i$ is assigned to machine $m$. The binary variables $y_{ij}$ are sequencing variables that are equal to one if jobs $i$ and $j$ are assigned to the same machine and job $j$ is processed after job $i$. Using the above described variables and parameters, the MILP model for job-shop scheduling is formulated as follows.

$$\min \sum_{i \in I} \sum_{m \in M} C_{im} x_{im} \tag{9}$$

$$s.t. \quad ts_i \geq R_i, \;\; \forall i \in I \tag{10}$$

$$ts_i \leq D_i - \sum_{m \in M} P_{im} x_{im}, \quad \forall i \in I \tag{11}$$

$$\sum_{m \in M} x_{im} = 1, \;\; \forall i \in I \tag{12}$$

$$y_{ij} + y_{ji} \geq x_{im} + x_{jm} - 1, \quad \forall i, j \in I, \; j > i, m \in M \tag{13}$$

$$ts_j \geq ts_i + \sum_{m \in M} P_{im} x_{im} - U(1 - y_{ij}), \qquad \forall i, j \in I, i \neq j \tag{14}$$

$$y_{ij} + y_{ji} \leq 1, \quad \forall i, j \in I, j > i \tag{15}$$

$$y_{ij} + y_{ji} + x_{im} + x_{jn} \leq 2, \qquad \forall i, j \in I, \ j > i, \forall m, n \in M, m \neq n \tag{16}$$

$$ts_i \geq 0, \ \forall i \in I; \quad x_{im} \in \{0,1\}, \ \forall i \in I, \forall m \in M; \quad y_{ij} \in \{0,1\}, \ \forall i, j \in I, i \neq j \tag{17}$$

The objective function in Eq. (9) is to minimize the processing costs associated with processing jobs assigned to the respective machines. Constraint (10) ensures that each job $i \in I$ is processed after its release date, and constraint (11) does not allow processing of any jobs later than their respective due dates. The assignment constraint (12) enforces that each job $i$ is processed by a single machine for this single-stage scheduling model. Constraint (13) models the logical relationship between the assignment variables and the sequencing variables. It implies that if jobs $i$ and $j$ are assigned to the same machine $m$, then the jobs must be processed one after the other. The parameter $U$ in sequencing constraint (14) is given by $U = \sum_{i \in I} max_{m \in M} \{P_{im}\}$. The sequencing constraint ensures that job $j$ starts processing after job $i$ finishes, provided that both jobs $i$ and $j$ are assigned to the same machine. Start times of both jobs remain independent of each other if they are assigned to different machines. Constraints (15) and (16) are simple logical cuts that reduce the computational time required to solve the MILP problem by a significant amount [59]. Constraint (15) is based on the logic relationship that either job $j$ is processed after job $i$ or vice versa, irrespective of their assigned machines. The last constraint ensures that the sequencing variables $y_{ij}$ and $y_{ji}$ are zero, if jobs $i$ and $j$ are assigned to different machines.

## 5.2. Hybrid QC-MILP Decomposition Method

The MILP problem for job-shop scheduling is difficult to solve by off-the-shelf optimization solvers due to the problem structure. Additionally, the sequencing constraints do not significantly tighten the LP relaxation of the problem [59]. The combinatorial nature stemming from mixed-integer terms leads to additional computational complexity. To address this computational challenge, we develop a hybrid solution strategy that integrates a decomposition-based algorithm with a QC solution technique for global optimization of this challenging job shop scheduling problem.

The main idea is that if a set of jobs cannot be scheduled on a particular machine, then it will be impossible to find a feasible schedule for any assignment that assigns all those jobs to that machine. The decomposition procedure selectively eliminates the possibility of such infeasible assignments by applying integer cuts. The proposed models and solution algorithm draw inspiration from the hybrid algorithm proposed by Jain and Grossmann [59]. The objective function (18) is the same as that of the original MILP model in Eq. (9). Constraints (10) - (12) and (17) form the constraints of the hybrid model described below. These timing and assignment constraints together form a new set of constraints (19) for the relaxed MILP model. Constraints (13) - (16) are concerned with sequencing jobs, and are reduced to the model in the QC step shown in (20). The QC step uses start times for each job as parameters in order to determine a sequence for the same. The Hamiltonian $H$ represents a single objective function, which uses the identical sequencing variables $y_{ij}$ and takes the form of a QUBO problem. The size of set $S$ in (20) depends on the number of jobs assigned to the same machine, for which scheduling start times have already been determined.

$$\min \sum_{i \in I} \sum_{m \in M} C_{im} x_{im} \tag{18}$$

$$\left. \begin{aligned} s.t.\ ts_i &\geq R_i \quad \forall i \in I \\ ts_i &\leq D_i - \sum_{m \in M} P_{im} x_{im}, \quad \forall i \in I \\ \sum_{m \in M} x_{im} &= 1, \quad \forall i \in I \\ ts_i &\geq 0, \quad \forall i \in I \\ x_{im} &\in \{0,1\}, \quad \forall i \in I, \forall m \in M \end{aligned} \right\} \text{Relaxed MILP} \tag{19}$$

$$\left. \begin{aligned} \min H &= \sum_{i,j \in S} 1 - y_{ij} - y_{ji} + 2 y_{ij} y_{ji} + y_{ij} \left( U \left( ts_i^* - ts_j^* \right) + \sum_{m \in M} P_{im} x_{im}^* \right) \\ S &= \left\{ i, j \mid x_{im}^* = x_{jm}^* = 1, \ \forall i,j \in I, \forall m \in M \right\} \end{aligned} \right\} \text{QC step} \tag{20}$$

The proposed hybrid QC-MILP decomposition method combines the deterministic aspect for solving the relaxed MILP problem with the quick search space traversal of QC techniques for solving the problem in the QC step. Details of this hybrid QC-MILP decomposition method for job-shop scheduling problem are presented in Figure 3. The algorithm described in this section has two phases. The first phase involves solving the relaxed MILP problem by the classical CPU-based deterministic Gurobi solver. Solutions to this relaxed MILP problem produces an assignment of machines to process each job, and is the partial optimal solution denoted by $x_{im}^*$ and $ts_i^*$. If the relaxed MILP problem is not feasible, then no solution exists for the original problem and the algorithms stops. The second phase is to determine a schedule for each machine and the assigned jobs using the QC step. Hamiltonian in the QC step uses the partial optimal solutions from the first phase, and is solved using the quantum processor in order to locate a feasible schedule in the integer space. As stated earlier, the size of this Hamiltonian is dependent on the number of machines, on which multiple jobs are assigned. An optimal solution to the job-shop scheduling problem can be returned only if a feasible schedule is obtained in the second phase of this decomposition algorithm. In case the QC step returns an infeasible solution, integer cuts are added to the relaxed MILP problem to exclude any conflicting assignments. The relaxed MILP problem is re-solved to obtain alternate assignment decisions that do not have such conflicting assignments from the previous step. If an alternate assignment is not possible, then the scheduling problem is infeasible. It is important to note that integer cuts are added cumulatively to ensure the success of the decomposition algorithm and that all feasibility checks are performed on a classical CPU-based computer.

The integer cuts added to the relaxed MILP problem after completion of phase two are critical to the decomposition algorithm. Integer cuts are only added for machines that could not be scheduled successfully. For each machine with an infeasible schedule, the integer cut formulation is $\sum_{i \in S'} x_{im} \leq |S'| - 1$, where $S'$ is the set of jobs assigned to machine $m$. For this parallel scheduling problem, the sequence for each machine could be determined separately. Instead, to avoid solving multiple QC problems associated with each machine, we choose to formulate the Hamiltonian in such a way that the sequence of all machines is determined together. Also, identifying the machines with an infeasible schedule becomes easier to detect and is done using a classical computer. Searching for feasible schedules at the QC step is an important aspect of the proposed hybrid QC-MILP decomposition method, and is crucial to obtain an optimal solution. The QC step could return multiple solutions due to QC's probabilistic nature, for which the feasibility of solutions is considered to continue the algorithm. Therefore, finding an optimal solution and proving optimality for the problem in the QC step can be difficult. If a feasible solution exists for the problem in the second phase of the hybrid algorithm, the algorithm will converge to a global optimum [59]. Thus, it should be noted that the hybrid QC-MILP decomposition method converges to an optimal solution or proves infeasibility in finite number of iterations. The decomposition scheme proposed by Jain and Grossman [59] plays an important role in solving the job-shop scheduling problem. The proposed hybrid QC-MILP decomposition method integrates QC-based solution techniques with the decomposition scheme to solve large-scale MILP problems and guarantee an optimal solution.

## 5.3. Computational Results

We carry out computational experiments for instances of the job-shop scheduling problem to illustrate the applicability of the proposed QC-MILP decomposition method. The job-shop scheduling problem being an MILP is solved using the MILP solver Gurobi on the classical computer mentioned in Section 3. The same classical computing configuration with Gurobi is used to solve the relaxed MILP problem of the hybrid QC-MILP decomposition method, to keep track of the iteration number, and to perform feasibility checks. The problem in QC step, on the other hand, is solved with the quantum processor reported in Section 3. Additionally, large Hamiltonians corresponding to the QC step were solved with the *qbsolv* utility tool with the quantum processor as backend. The sub-QUBOs are compiled on the same quantum processor. Both classical and quantum computation times are recorded at each iteration of the QC-MILP decomposition method, and the total computation times are reported for each instance. This time is the wall-clock time required by the hybrid QC-MILP decomposition method to return an optimal solution. For comparison, the objective function values and computational times required to solve the original MILP problems directly with Gurobi are also reported. Table 3 shows these details from computational experiments using different solvers and algorithms for the job-shop scheduling instances.

Problems of various sizes are considered in this section. The size of the job-shop scheduling problem depends on the number of jobs to be processed and the available machines. Randomly generated instances ranging from 40 jobs and machines to 150 jobs and machines are solved to identify the varying trends of computation time and solutions quality. Problem sizes of these instances are also reported in Table 3. For the largest problem with 150 jobs and 150 machines, the original MILP problem contains 150 continuous variables, 44,850 binary variables and 126,579,825 constraints. Note that job-shop scheduling problem size increases quadratically with the number of jobs and machines.

---

**Hybrid QC-MILP decomposition method**

---

1:  $iter \leftarrow 0, Opt \leftarrow$ False

2:  **while** $Opt$ is False **do**

3:  $\quad iter \leftarrow iter + 1$

4:  $\quad$ Solve relaxed MILP problem in (19) to minimize $\sum_{i \in I} \sum_{m \in M} C_{im} x_{im}$.

5:  $\quad$ **if** relaxed MILP problem is infeasible

6:  $\quad\quad$ No solution exists.

7:  $\quad\quad$ Stop procedure and exit the loop.

8:  $\quad$ **else**

9:  $\quad\quad$ Return partial optimal solution $x_{im}^*, ts_i^*$.

10: $\quad$ **end if**

11: $\quad$ Solve problem in QC step to determine schedule with Hamiltonian given in Eq. (20).

12: $\quad$ **if** feasible schedule not found

13: $\quad\quad$ Identify machines for which feasible schedule could not be found.

14: $\quad\quad$ Add respective integer cuts to the relaxed MILP problem in (19).

15: $\quad\quad$ $\displaystyle\sum_{i \in S'} x_{im} \leq |S'| - 1$

$\quad\quad$ where $S' = \{i \mid x_{im} = 1, \forall i \in I, \forall m \in M\}$

16: $\quad\quad$ $Opt \leftarrow$ False

17: $\quad$ **else**

18: $\quad\quad$ Return optimal solution $x_{im}^*, ts_i^*, y_{ij}^*$.

19: $\quad\quad$ $Opt \leftarrow$ True

20: $\quad$ **end if**

21: **end while**

22: **return** $x_{im}^*, ts_i^*, y_{ij}^*$

---

*Figure 3. Hybrid QC-MILP decomposition method for solving job-shop scheduling problem*

The size of QUBO problem in the QC step, however, changes dynamically with each iteration of the decomposition procedure and depends on the optimal assignments in the previous step. In order to illustrate the applicability of the proposed solution strategy, a small scheduling instance comprising of 8 jobs to be scheduled on 8 parallel machines is also considered. This instance was solved using Gurobi solver and the hybrid QC-MILP decomposition technique. Figure 4 represents the obtained schedule in each case. As seen in the chart in Figure 4a, multiple jobs have been scheduled on two single machines. Specifically, jobs 7 and 8 have been scheduled on machine 7, with jobs 3 and 4 scheduled on machine 2. Unlike the schedule obtained with Gurobi solver, the hybrid QC-MILP decomposition method yields an alternative schedule that utilizes each available machine to process the jobs. The alternative optimal schedule is shown in Figure 4b.

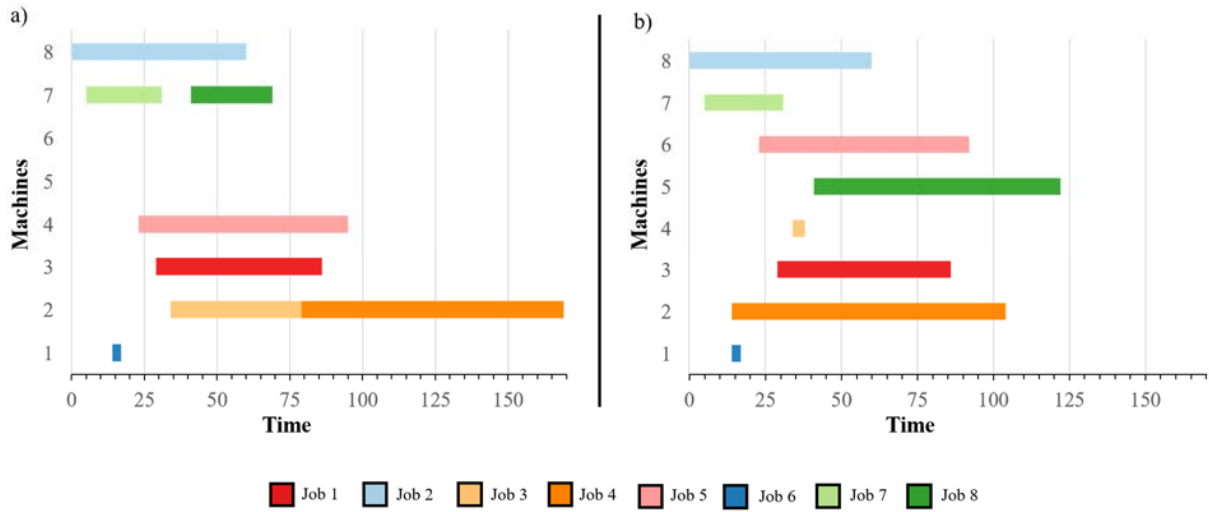

*Figure 4. Gantt charts of scheduling results of a job-shop scheduling problem with eight jobs and machines obtained using a) MILP solver Gurobi and b) Hybrid QC-MILP decomposition method.*

Comparing the computational times and objective values reported in Table 2, we can see that the hybrid QC-MILP decomposition method performs competitively against the Gurobi solver. General-purpose solvers like Gurobi are developed from the ground up to exploit modern classical computing architectures for solving MILP problems. Therefore, small job-shop scheduling problems can be solved within short computation times by these state-of-the-art classical solvers. With Gurobi solver, the size of the branch-and-bound tree exceeds the allotted physical memory and does not return any solution for job-shop scheduling instances with more than 70 jobs and machines. Yet, the proposed hybrid QC-MILP decomposition strategy yields a solution for each of these instances within reasonable computational time without any memory limitations. Iterations required for convergence to optimal solution in the decomposition procedure increase with the problem size. It can also be seen that classical resource utilization time, which is the time taken by Gurobi solver to solve the relaxed MILP problems, is much less compared to the computation time utilized by the quantum processor. However, it should be noted that quantum time does not only represent quantum annealing time exclusively, but also includes time required to partition the large QUBO problems in the QC step. Increasing the problem size increases the number of assignments in the first phase of the algorithm. As scheduling the jobs on machines is a demanding task, large-scale problems require longer computation time to determine a feasible schedule in the second phase. Integer cuts added to the original constraints in the relaxed MILP problem also contribute to the efficiency of the decomposition algorithm.

The disadvantage of the original MILP model for the job-shop scheduling problem is that the sequencing constraints do not contribute directly to the objective function value. Alternatively, the hybrid QC-MILP decomposition method includes only assignment constraints in the relaxed MILP model, while the QC step effectively uses the quantum annealing process to determine a feasible schedule. Clearly, the relaxed MILP problem is smaller than its corresponding original MILP problem and can be solved with less computation time. It can be argued that memory limitation of the Gurobi solver can be overcome by using classical computers of larger physical memory. However, as the branch and bound tree size increases exponentially with problem size, the MILP solver does not guarantee a solution for larger job-shop scheduling instances. Advantages of increasing allotted physical memory are disproportionate to the expected improvement in problem size solvable by Gurobi. From a practical standpoint, 150 jobs and 150 machines is a reasonable industrial size problem. The hybrid QC-MILP decomposition method yields an optimal solution without any physical memory augmentation at the cost of reasonably longer runtimes. We note that a global optimal solution is guaranteed by the hybrid QC-MILP decomposition method. This job-shop scheduling application demonstrates that the complementary strengths of MILP and QC methods are able to tackle more complex and large-scale problems.

*Table 3. Computational results of the job-shop scheduling problems*

| Jobs and Machines | Continuous variables | Binary variables | Gurobi 8 | | Hybrid QC-MILP decomposition method | | | | |
|---|---|---|---|---|---|---|---|---|---|
| | | | Time (s) | Min obj. | Iterations | Time (s) | | | Min obj. |
| | | | | | | Classical | Quantum | Total | |
| 50 | 50 | 4,950 | 79 | 50.0 | 142 | 51 | 177 | 228 | 50.0 |
| 60 | 60 | 7,140 | 243 | 60.0 | 169 | 78 | 555 | 633 | 60.0 |
| 70 | 70 | 9,730 | 405 | 71.0 | 270 | 162 | 1,886 | 2,048 | 71.0 |
| 80 | 80 | 12,720 | -- | --* | 299 | 212 | 5,173 | 5,385 | 80.0 |
| 90 | 90 | 16,110 | -- | --* | 367 | 342 | 9,504 | 9,846 | 90.0 |
| 100 | 100 | 19,900 | -- | --* | 552 | 3,630 | 23,176 | 26,806 | 101.0 |
| 110 | 110 | 24,090 | -- | --* | 666 | 936 | 22,287 | 23,223 | 110.0 |
| 120 | 120 | 28,680 | -- | --* | 775 | 1,355 | 38,141 | 39,496 | 120.0 |
| 130 | 130 | 33,670 | -- | --* | 805 | 1,549 | 35,213 | 36,762 | 130.0 |
| 150 | 150 | 44,850 | -- | --* | 1,252 | 3,684 | 61,244 | 64,928 | 150.0 |

* Memory limit of 32 GB RAM exceeded and no solution is returned.

# 6. QC for Manufacturing Cell Formation

Cellular manufacturing is an important application of group technology and is being widely applied in manufacturing industries [44]. In this manufacturing approach, the equipment is arranged to facilitate continuous flow production, resulting in increased work flow, reduced response and production times, and increased profits. The basic idea underlying cellular manufacturing is to divide the manufacturing system into several cells. Similar parts are processed in the same cell, such that the interactions of machines and parts within a cell are maximized to improve efficiency. The first step of cellular manufacturing system design is cell formation which involves selecting parts and machines that will be allocated to each cell [60]. The objective of the manufacturing cell formation problem is to minimize the total cost associated with intracellular movement, resource utilization, and machine set-ups. Manufacturing cell formation belongs to the class of complex NP-hard optimization problems and has received significant attention. Metaheuristic techniques have demonstrated exceptional performance for solving some variants of the manufacturing cell formation problem, but they can only obtain near-optimal solutions for small to medium-sized problems [61]. Exact solution methods like branch-and-cut consume a considerable amount of time to obtain a global optimum for large-scale manufacturing cell formation problems [61]. Hybridization of exact and metaheuristic algorithms has proven to be a more viable option that uses complementary strengths of both techniques, and overcomes difficulties associated with determining mutually separable cells [62]. Such hybrid techniques can obtain optimal solutions in most cases and solve large problems with satisfactory results.

## 6.1. Model Formulation

Several variants of the manufacturing cell formation model exist in literature. Factors like resource and operational costs, intra-cellular movement costs, resource utilization costs, grouping efficiency, and others are considered in the manufacturing cell formation model formulation. Here we consider the MIQP model where the operational requirements for each part are known beforehand. This formulation involves grouping a set of parts $P$ and a set of machines $M$ into subsystems termed as cells denoted by the set $R$. The cost of inter-cell movement per unit of part $i \in P$ is given by $c_i$, with $v_i$ units of each part $i$ in the manufacturing system. The cost of part $i \in P$ not utilizing machine $j \in M$ is represented by $u_{ij}$. Each part $i$ needs to be processed $o_{ij}$ times on machine $j$, and $a_{ij}$ is a real-valued parameter that indicates whether a part $i$ requires machine $j$. A non-zero value of $a_{ij}$ implies that part $i$ requires machine $j$ for processing. These operational parameters for cellular manufacturing are known a priori and remain independent of any external factors like product demand and operational changes.

The major decisions involved in the manufacturing cell formation problem is to determine the parts and machines assigned to each cell $k \in R$. The continuous assignment variables $x_{ik}$ denote whether part $i$ is assigned to cell $k$, and binary assignment variable $y_{jk}$ is equal to one when machine $j$ is assigned to the cell $k$. Assignment variable $x_{ik}$ is bounded between zero and one with its non-zero value implying that part $i$ is processed in cell $k$. Using the above described variables and parameters, the MIQP model for the manufacturing cell formation problem can be formulated as follows.

$$\min \sum_{i \in P} \sum_{j \in M} \sum_{k \in R} c_i v_i o_{ij} a_{ij} x_{ik} \left(1 - y_{jk}\right) + \sum_{i \in P} \sum_{j \in M} \sum_{k \in R} u_{ij} v_i \left(1 - a_{ij}\right) x_{ik} y_{jk} \tag{21}$$

$$s.t. \quad \sum_{k \in R} x_{ik} = 1, \quad \forall i \in P \tag{22}$$

$$\sum_{k \in R} y_{jk} = 1, \quad \forall j \in M \tag{23}$$

$$0 \le x_{ik} \le 1, \quad \forall i \in P, \forall k \in R \tag{24}$$

$$y_{jk} \in \{0,1\}, \quad \forall j \in M, \forall k \in R \tag{25}$$

The objective function in (21) represents the total cost to be minimized where the first term represents the total cost of inter-cell movement. Second term of this objective function represents the total cost of resource underutilization. It should be noted that the variable $x_{ik}$ and parameter $a_{ij}$ are set as real numbers bounded between [0,1] to ensure consideration of alternate routings. Constraint (22) ensures allocation of each part to a cell. Similarly, constraint (23) ensures that each machine can be assigned to only one cell. In this MIQP model, we do not place any restrictions on machine pairs in a particular cell, and it is assumed that any machine can be placed in any cell irrespective of other assignments. Capacity limitations of the number of machines in each cell are also discarded to allow flexibility, but such restrictions can be easily considered by adding constraints of the form $M_{min} \le \sum_{j \in M} y_{jk} \le M_{max}$. It should be noted that empty cells are also allowed in the formulated model.

## 6.2. Hybrid QC-MIQP Stepwise Decomposition Method

MIQP optimization problems are frequently considered expensive to solve. There are several methods available to efficiently solve such problems. The branch-and-cut method and generalized Bender's decomposition based methods are some of them [63, 64]. However, as the size of the MIQP problem grows, it is difficult to handle by off-the-shelf optimization solvers directly. Additionally, due to the combinatorial nature and nonlinearity stemming from the quadratic objective function, large-scale MIQP problems could be challenging to solve. To tackle this computational challenge, we propose a novel hybrid solution strategy to solve the manufacturing cell formation problem.

The proposed hybrid QC-MIQP stepwise decomposition method is based on Bender's decomposition [65], by considering the above manufacturing cell formation problem given in Eq. (21) to (24) as the primal problem. The dual of this problem is constructed after introducing new variables to replace the quadratic terms in the primal problem. It is important to note that we do not need dual variables corresponding to the upper bound in constraint (24). The main idea behind the decomposition algorithm is to iteratively generate upper and lower bounds on the optimal value by solving smaller subproblems.

The primal problem is MIQP with quadratic terms in the objective function using the set of real variables $x_{ik}$ and binary variables $y_{jk}$. The dual problem is constructed corresponding to this primal problem, and is referred to as the dual LP model. The dual LP model is linear with objective function (26) and constraints given in (27). This problem consists of four sets of real variables $l_{ijk}$, $m_{ijk}$, $n_{ijk}$ and $s_i$, where $l_{ijk}$, $m_{ijk}$ and $n_{ijk}$ are nonnegative variables and $s_i$ are free variables.

$$\max \sum_{i \in P} s_i - \sum_{i \in P} \sum_{j \in M} \sum_{k \in R} y_{jk}^* m_{ijk} + \sum_{i \in P} \sum_{j \in M} \sum_{k \in R} \left( y_{jk}^* - 1 \right) n_{ijk} \qquad (26)$$

$$s.t. \quad \sum_{j \in M} l_{ijk} - \sum_{j \in M} n_{ijk} + s_i \leq \sum_{j \in M} c_i v_i a_{ij} o_{ij}, \quad \forall i \in P, \forall k \in R$$
$$\left.\begin{array}{l} n_{ijk} - l_{ijk} - m_{ijk} \leq u_{ij} v_i \left(1 - a_{ij}\right) - c_i v_i o_{ij} a_{ij}, \quad \forall i \in P, \forall j \in M, \forall k \in R \\ l_{ijk}, m_{ijk}, n_{ijk} \geq 0 \\ s_i \text{ unbounded} \end{array}\right\} \text{Dual LP} \qquad (27)$$

Optimal values of the variables $m_{ijk}$, $n_{ijk}$ and $s_i$, obtained by solving the dual LP problem are used as parameters in formulating the model described in the QC step and are denoted as $\hat{m}_{ijk}, \hat{n}_{ijk}$, and $\hat{s}_i$, respectively. The objective of the QC step in (28) is to determine the assignments of machines to respective cells. The Hamiltonian $H$ represents a single objective function that uses the sequencing variables $y_{jk}$ and takes the form of a QUBO problem. It comprises of two distinct Hamiltonians represented by $H_{obj}$ and $H_c$ corresponding to the objective function and constraint (23), respectively. Parameter values $A$ and $B$ are fixed and determined empirically. QC step is dynamic in nature and changes with each solution iteration. Additionally, size of the problem in the QC step that contains all possible machine-cell assignments, remains constant.

$$
\left.
\begin{aligned}
&\min H = H_{obj} + H_c \\
&s.t.\ H_{obj} = -\sum_{j \in M} \sum_{k \in R} \left( \sum_{t=1}^{T} A Q_{jkt} + 2 B Q_{jkt} \left( F_t - Z_{T-1}^* \right) \right) y_{jk} \\
&\qquad\quad + \sum_{j \in M} \sum_{k \in R} \sum_{m \in M} \sum_{n \in R} \left( \sum_{t=1}^{T} Q_{jkt} Q_{mnt} \right) y_{jk} y_{mn} \\
&\quad\ H_c = \sum_{j \in M} \left( 1 - \sum_{k \in R} y_{jk} + 2 \sum_{k \in R} \sum_{n > k, n \in R} y_{jk} y_{jn} \right) \\
&\quad\ Q_{jkt} = \sum_{i \in P} \hat{m}_{ijk} - \hat{n}_{ijk} \\
&\quad\ F_t = \sum_{i \in P} \hat{s}_i - \sum_{i \in P} \sum_{j \in M} \sum_{k \in R} \hat{n}_{ijk}
\end{aligned}
\right\} \text{QC step} \qquad (28)
$$

The proposed hybrid QC-MIQP stepwise decomposition method combines the deterministic aspect of dual LP model with quantum approach of the QC step and yields an optimal solution for the manufacturing cell formation problem. Details of this algorithm are shown in Figure 5. An outer loop for this algorithm keeps a record of the lower bound $LB$, upper bound $UB$ and the iterations $T$. First step of the hybrid QC-MIQP stepwise decomposition technique involves initializing all assignment variables $y_{jk}$, $LB$ and $UB$, as shown in the figure. Optimal values of the dual variables are obtained through solving the dual LP problem using a CPU-based classical computer. An infeasible dual LP problem implies that the original manufacturing cell formation problem is unbounded. The upper bound is updated based on the objective function value of the dual LP problem denoted by $Z_T^*$. The problem in the QC step is constructed using the optimal dual variable values, and is solved on a quantum processor to yield the assignment decisions $\hat{y}_{jk}$ corresponding to all machine-cell pairs. The objective value obtained through the QC step, $\hat{Z}_T$ during the $T^{th}$ decomposition iteration, is determined using equation $\hat{Z}_T = \max_t (F_t - \sum_{j \in M, k \in R} Q_{jkt} \hat{y}_{jk})$. Functional forms of $F_t$ and $Q_{jkt}$ are provided in the QC step. The lower bound is updated using this maximum value obtained through solving the problem in the QC step. The algorithm stops if the lower bound assumes a value higher than that of the upper bound, with the upper bound as the optimal solution of the manufacturing cell formation problem. Alternatively, if the lower bound value is less than the upper bound value, assignment variables used in the dual LP problem are updated to those obtained from solving the problem in the QC step, and the same procedure is repeated. It should be noted that the feasibility checks, as well as value comparisons, were performed on a classical CPU-based computer. Feasibility of the dual LP problem is determined by the MILP solver Gurobi. This solution algorithm converges within finite iterations. The probabilistic nature of the solutions returned at the QC step might affect the total computational time, by requiring fewer or more iterations. A global optimum might not be guaranteed by the proposed hybrid QC-MIQP stepwise decomposition method, but varying the Hamiltonian in the QC step can eliminate its heuristic nature. Studying the impacts of the Hamiltonian on the convergence of this decomposition algorithm is beyond the scope of this work. The manufacturing cell formation problems are solved using the proposed hybrid QC-MIQP stepwise decomposition method that is based on generalized Bender's decomposition [65], which has been used in batch manufacturing systems [66, 67]. The upper bound computed by solving the dual LP problem does not need to be a global optimum, resulting in a trade-off between the number of iterations until convergence and the required computation time.

## 6.3. Computational Results

In order to illustrate the applicability of the proposed solution strategy, we carry out computational experiments on several manufacturing cell formation instances. The formulated manufacturing cell formation problem is a MIQP problem and can be solved using the Gurobi solver. A time limit of 24 hours is enforced on the Gurobi solver for an appropriate comparison with the proposed hybrid QC-based solution strategy. The classical computer reported in Section 3 is used to solve the MIQP problems and as a backend to solve the dual LP problem, as well as to perform feasibility checks for the proposed hybrid QC-MIQP stepwise decomposition method. The quantum computer mentioned in Section 3, on the other hand, is used to solve the problem in the QC step. Large Hamiltonians in the QC step are solved with *qbsolv* utility tool and the same quantum processor as backend. At each iteration of the decomposition procedure, computation times for each of classical and quantum backend are recorded. The total computation times for the hybrid QC-MIQP stepwise decomposition method and the original MIQP problem solved with Gurobi, along with the obtained objective function values, are reported in Table 4 for each of the manufacturing cell formation instances.

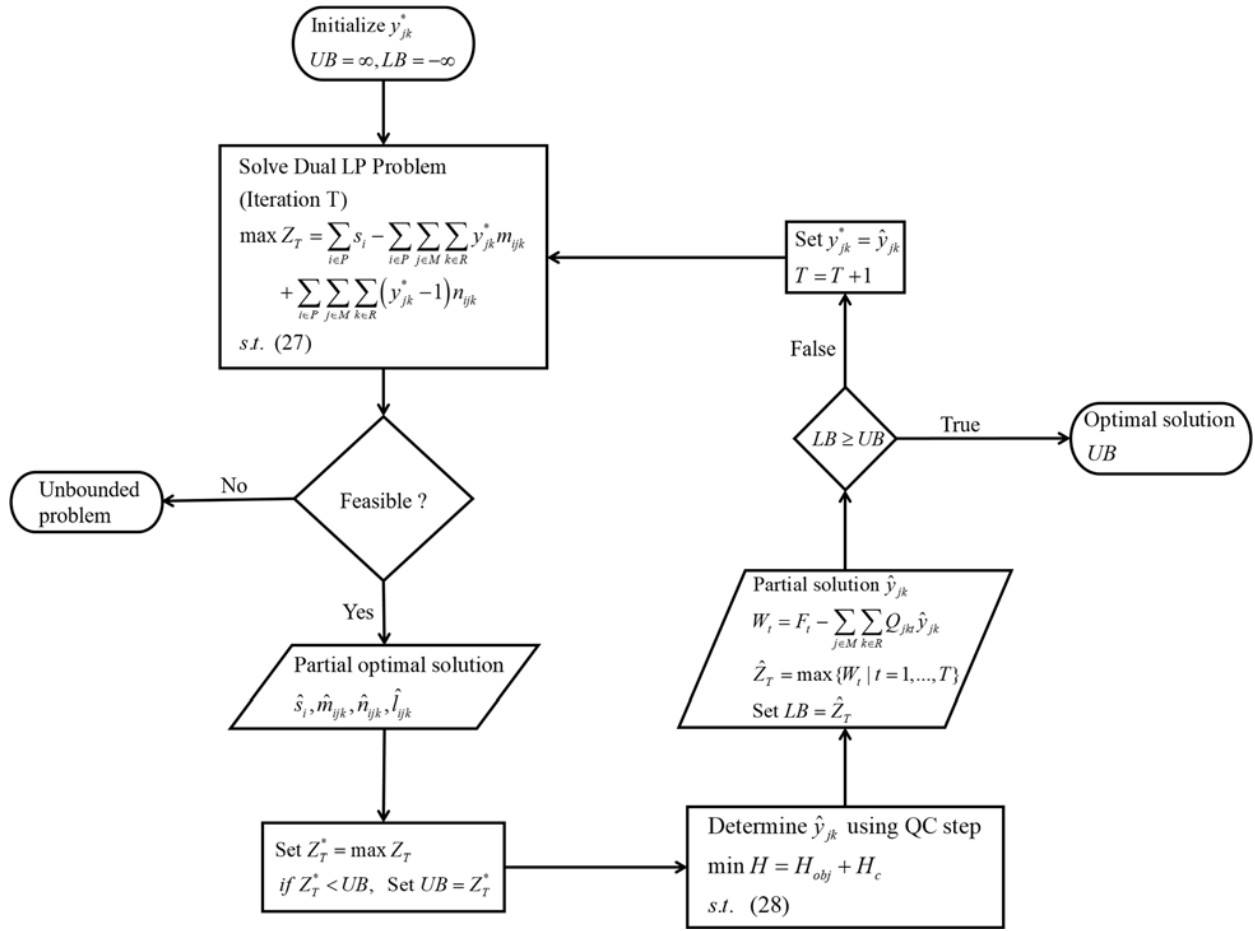

*Figure 5. Hybrid QC-MIQP stepwise decomposition method for solving the manufacturing cell formation problem*

In this section, we consider manufacturing cell formation problems of various sizes. The instances are chosen such that their problem sizes increase gradually in terms of both continuous and binary variables. It is also ensured that a feasible solution exists for each of these randomly generated instances. This is achieved by fixing the number of available machines that satisfy capacity constraints to the ratio of total time required for all processing operations and the total available operating time. The size of manufacturing cell formation problem depends on the number of types of parts to be processed, number of machines available and the number of cells permitted. Problem sizes for the used manufacturing cell formation instances are also reported in Table 4. The largest instance solved with 75 parts, 70 machines, and 15 cells comprises of 1,050 binary variables, 1,125 continuous variables and 1,270 constraints in the MIQP problem. Size of this MIQP problem increases with the product of the numbers of parts and cells, and the number of machines and cells as well. It is also interesting to note that the size of QUBO problem in the QC step is fixed throughout the decomposition procedure and comprises of an equal number of binary variables as the original MIQP problem. We also consider a small instance with five parts and machines to be divided into two cells to illustrate the application of the manufacturing cell formation problem in Figure 6. The interactions between machines and parts processed by them are represented by lines in this figure. The objective of manufacturing cell formation problem is to arrange these parts and machines into cells such that inter-cell movement cost along with resource under-utilization is minimized. In this case, formation of cells results in no inter-cell movement and is represented by Figure 6b.

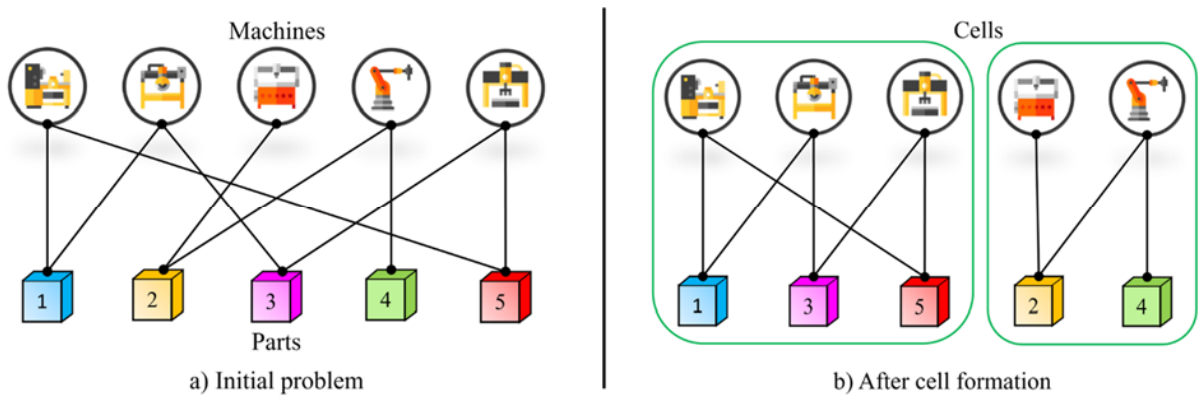

*Figure 6. Interactions between the machines and the processed parts a) before cell formation, and b) after cell formation.*

From the computational results presented in Table 4, it can be clearly seen that the proposed hybrid QC-MIQP stepwise decomposition method performs more efficiently than the conventional Gurobi solver in terms of both solution quality and computation time for medium to large size problems. The mixed-integer programming solver Gurobi utilizes an advanced branch-and-cut algorithm that can quickly and robustly solve MIQP problems. The MIQP problems with fewer variables and constraints, can therefore be solved much more efficiently than the proposed hybrid QC-MIQP stepwise decomposition method. The performance of Gurobi solver deteriorates beyond instances with 500 binary variables, and a clear quantum advantage is perceived with the hybrid QC-MIQP stepwise decomposition method. There were no specific trends observed for the number of iterations required for convergence to the optimal solution, in the decomposition procedure. It can also be observed that larger manufacturing cell formation instances could not be solved to optimality by the state-of-the-art mixed-integer programming solver Gurobi, while the proposed hybrid QC-MIQP stepwise decomposition method obtains optimal or near-optimal solutions with almost half of that time. Although classical resource utilization time is much less compared to the quantum time, it should be noted that quantum time does not only represent annealing time exclusively, but also includes the time required to partition the large QUBO problems in the QC step.

Although the MIQP problem of manufacturing cell formation is loosely constrained, the main disadvantage faced when solving this problem with mixed-integer programming solver is the quadratic nature of its objective function. Size of the dual LP problem is much larger than that of its quadratic variant, but this problem can be solved with ease by a linear solver. As mentioned earlier, problem size in the QC step remains constant throughout the decomposition procedure, and hence the corresponding QUBO problem can be efficiently solved after an embedding scheme is determined for this QUBO problem by simply changing edge and node weights. The heuristic nature of QC techniques when combined with the deterministic aspect of solvers like Gurobi can prove beneficial, and is demonstrated through this manufacturing cell formation application.

*Table 4. Computational results of the manufacturing cell formation problems*

| Parts | Machines | Cells | Continuous variables | Binary variables | Gurobi 8 | | Hybrid QC-MIQP stepwise decomposition method | | | | |
|---|---|---|---|---|---|---|---|---|---|---|---|
| | | | | | Time (s) | Min obj. | Iterations | Time (s) | | | Min obj. |
| | | | | | | | | Classical | Quantum | Total | |
| 10 | 10 | 4 | 40 | 40 | 0.09 | 771 | 25 | 5 | 0.97 | 6.38 | 771 |
| 20 | 25 | 5 | 100 | 125 | 6 | 6,467 | 20 | 7 | 13,747 | 13,754 | 6,467 |
| 25 | 30 | 7 | 175 | 210 | 8 | 9,197 | 24 | 36 | 32,540 | 32,566 | 9,243 |
| 35 | 35 | 8 | 280 | 280 | 180 | 16,130 | 29 | 30 | 25,827 | 25,857 | 16,130 |
| 45 | 40 | 9 | 405 | 360 | 36,006 | 18,913 | 25 | 42 | 22,854 | 22,896 | 18,913 |
| 45 | 45 | 9 | 405 | 405 | 17,322 | 26,573 | 29 | 54 | 24,164 | 24,218 | 26,753 |
| 50 | 50 | 10 | 500 | 500 | 75,448 | 32,065 | 26 | 83 | 38,819 | 38,902 | 32,065 |
| 50 | 60 | 10 | 500 | 600 | 86,400* | 37,105 [a] | 29 | 116 | 46,316 | 46,432 | 37,105 |
| 75 | 60 | 15 | 1,125 | 900 | 86,400* | 55,478 [a] | 27 | 178 | 45,311 | 45,489 | 55,478 |
| 75 | 70 | 15 | 1,125 | 1,050 | 86,400* | 59,888 [a] | 31 | 253 | 75,409 | 75,662 | 59,888 |

* Timeout of 24 hours (86,400 CPUs) reached and the best solution found is reported.

[a] The optimality gap for best solution found is more than 34% after running for 24 hours.

# 7. QC for Vehicle Routing

An important logistics optimization problem is the vehicle routing problem. This problem is concerned with determining an optimal set of routes for a fleet of vehicles in order to serve a given set of customers or locations, in order to minimize the total transportation cost [68]. Operational constraints must also be satisfied when minimizing the global transportation cost. Due to the versatile nature and richness in terms of real-world applicability, the vehicle routing problem has attracted significant attention [69]. In addition, the vehicle routing problem leads to challenging formulations that belong to the NP-hard computational complexity class and requires development of sophisticated solution strategies [70]. Sophisticated exact solution techniques based on integer programming like branch-and-cut and branch-and-prize can solve medium-sized vehicle routing problems, but success of these methods depend on the model formulation [71]. As most of the real-world applications consist of hundreds of customers or locations to be serviced, the focus has largely been on the development of approximate solution techniques that can provide high-quality solutions within reasonable computation time. Heuristic and metaheuristic algorithms can be directly applied to several variants of the vehicle routing problem [41, 72], but such techniques are context dependent and require careful parameter tuning to obtain good-quality solutions [73, 74]. Hybrid techniques can combine different components of both exact and heuristic search schemes to overcome such difficulties.

## 7.1. Model Formulation

Mathematical programming formulations for the vehicle routing problems can be broadly classified into two categories, namely the vehicle flow formulation and the set partitioning formulation [75]. The vehicle flow formulation leads to a compact model, while the set partitioning formulation has a large number of variables to represent all possible routes but fewer constraints. Here we consider the quadratic vehicle flow model for a variant of the capacitated vehicle routing problem. The purpose of this quadratic formulation is to significantly reduce the number of constraints, so that it can also serve as a basis for more complicated vehicle routing problem variants [76]. This formulation involves a vertex set $V$ representing the customer and depot locations, where the vertex 0 represents the depot. Customers are to be serviced by the available vehicles in set $H$. Set $N$ is a collection of steps covered by the vehicle. The maximum number of allowable steps for a vehicle is equal to the number of vertices so as to avoid multiple visits to the same location. The cost of travelling from location $i \in V$ to location $j \in V$ is denoted by $C_{ij}$, and the working time for the same pair of locations is represented by $W_{ij}$. The working time and travel costs between two locations are not considered to be proportional as the costs and time are influenced by several external factors, thus mimicking real-world conditions.

The vehicle routing problem aims to determine at most $|H|$ optimal routes such that specific design requirements are satisfied along with operational constraints, while minimizing total travel cost/time or maximizing profits. The decision variables in this model are the set of binary variables $x_{ip}^v$ that indicate whether the customer location $i \in V$ is visited by vehicle $v \in H$ at step $p \in N$ of its route. Using the above described variables and parameters, the vehicle flow model for the vehicle routing problem can be written as follows.

$$\min \frac{\sum_{v \in H} \sum_{i \in V} \sum_{j \in V} \sum_{p \in N} C_{ij} x_{ip}^v x_{jp+1}^v}{\sum_{v \in H} \sum_{i \in V} \sum_{j \in V} \sum_{p \in N} W_{ij} x_{ip}^v x_{jp+1}^v} \tag{29}$$

$$s.t. \sum_{v \in H} \sum_{p \in N} x_{ip}^v = 1, \quad \forall i \in V \setminus \{0\} \tag{30}$$

$$\sum_{i \in V \setminus \{0\}} x_{ip}^v = \sum_{i \in V} x_{ip+1}^v, \quad \forall p \in N, \forall v \in H \tag{31}$$

$$x_{ip}^v \in \{0,1\}, \quad \forall v \in H, \forall p \in N, \forall i \in V \tag{32}$$

The objective of this problem in (29) is to minimize the logistic ratio defined as the ratio of total cost incurred to the overall resources spent to serve the customers. Such fractional objective function formulations have been previously applied to real-world large-scale routing problems with inventory management [77]. The numerator of objective function in this IQFP model indicates the total travelling cost of vehicles; the denominator represents the total working time used to serve all the customer locations. Each customer must be visited and serviced only once by exactly one vehicle, as given by constraint (30). The set $V \setminus \{0\}$ represents all the customer locations only, where the $0^{th}$ vertex is the depot location. Constraint (31) enforces that a vehicle servicing a customer must leave for another customer or return to the depot in the next step of its route. We do not place any capacity restrictions on the vehicle, and assume complete flexibility of its demand and supply operations. The demand capacity constraints are not considered in this vehicle flow model. However, such constraints can be easily incorporated into this formulation to facilitate the formulation of some complex variants of the vehicle routing problem.

## 7.2. Hybrid QC-IQFP Parametric method

As mentioned above, the vehicle routing problem is an IQFP problem that involves a fractional objective function. Notably, fractional programs have been known to be intrinsically difficult to optimize globally [78]. Because of its combinatorial nature and pseudo-convexity, the IQFP problem can be computationally intractable [79]. Moreover, the quadratic terms in the fractional objective function adds to the complexity of this problem, resulting in a challenging optimization problem that could be difficult to handle by off-the-shelf MINLP solvers directly. To tackle this computationally challenging problem, we develop a hybrid algorithm which adopts an efficient parametric algorithm [80] along with sophisticated QC-based techniques.

The vehicle routing problem given in Eq. (29)-(32) can be solved using the proposed hybrid QC-IQFP parametric method. This method uses an extension of inexact parametric algorithm as a basis framework for the global optimization of fractional programming problems [80]. The proposed hybrid parametric method revolves around the idea of iteratively solving the problem in the QC step until convergence is achieved. The objective of QC step in (33) is to determine the set of optimal routes for the formulated vehicle routing problem. Solving the problem in the QC step minimizes the Hamiltonian $H$ that takes the form of a QUBO problem. This Hamiltonian comprises of two separate Hamiltonians $H_{obj}$ and $H_c$ shown in the QC step that correspond to the quadratic objective function and the route and service constraints, respectively. The model in the QC step uses a parameter $\lambda$ that is dynamic in nature and changes with each iteration. Penalty weight $A$ in the QUBO problem also changes with each iteration and is set considerably higher than any coefficient in $H_{obj}$. Thus, the QUBO problem in the QC step is dynamic in nature, but the size of this QUBO problem remains constant with the number of binary variables equal to that of variables in the original IQFP problem.

$$\min H = H_{obj} + AH_c$$

$$s.t.\ H_{obj} = \sum_{v \in H} \sum_{i \in V} \sum_{j \in V} \sum_{p \in N} \left( C_{ij} - \lambda W_{ij} \right) x_{ip}^v x_{jp+1}^v$$

$$H_c = \sum_{i \in V \setminus \{0\}} \left( \sum_{v \in H} \sum_{p \in N} x_{ip}^v - 1 \right)^2 + \sum_{v \in H} \sum_{p \in N} \left( \sum_{i \in V \setminus \{0\}} x_{ip}^v - \sum_{i \in V} x_{ip+1}^v \right)^2 \quad \text{QC step} \qquad (33)$$

The proposed QC-IQFP parametric method exploits the deterministic aspect of inexact parametric algorithm and quantum search to obtain solutions for the quadratic objective function. Solution obtained by solving the QUBO problem in the QC step lie within the feasible search space of the original IQFP vehicle routing problem. Details of this hybrid parametric algorithm are provided in Figure 7 below.

**Hybrid QC-IQFP parametric method**

1: $iter \leftarrow 0, \lambda \leftarrow 0, obj \leftarrow +\infty, \delta \leftarrow 10^{-6}$

2: **while** $|\lambda - obj| > \delta$, **do**

3:     $iter \leftarrow iter + 1$

4:     Solve the Hamiltonian $H$ in QC step given by Eq. (33).

5:     **if** infeasible solution is returned

6:         No solution exists.

7:         Stop procedure and exit the loop.

8:     **else**

9:         Return partial optimal solution $\hat{x}_{ip}^v$

10:     **end if**

11:     $obj \leftarrow \lambda$

12:     $\lambda \leftarrow \dfrac{\sum_{v,i,j,p} C_{ij} \hat{x}_{ip}^v \hat{x}_{jp+1}^v}{\sum_{v,i,j,p} W_{ij} \hat{x}_{ip}^v \hat{x}_{jp+1}^v}$

21: **end while**

22: **return** $\hat{x}_{ip}^v, obj$

*Figure 7. Hybrid QC-IQFP parametric method for solving vehicle routing problem*

The parametric solution strategy is initialized by setting the parameter λ to zero and a user-defined tolerance value. An outer loop of this algorithm keeps track of the parameter values and the number of iterations. During each iteration of the process, problem in the QC step is solved using a quantum processor to obtain a set of feasible vehicle routes denoted by $\hat{x}_{ip}^v$. Using this set of routes, the objective function value is computed for the original IQFP problem given in (29). The absolute difference between this fractional objective function and the parameter λ is used to determine the convergence criteria. For the algorithm to stop iterating, this absolute difference should be less than or equal to the pre-defined tolerance value. Upon reaching convergence, the partial solution $\hat{x}_{ip}^v$ is returned as an optimal solution to the vehicle routing problem. The algorithm continues if convergence criteria is not met, and the problem parameters λ and $A$ in the QC step are updated. The parameter λ is updated to assume a new value equal to that of previously computed fractional objective function value. The problem in the QC step with updated parameter values is solved repeatedly until an optimal solution for the vehicle routing problem is found. It should be noted that programming functions like updating parameter values and checking convergence criteria are performed on a CPU-based classical computer. The inexact parametric method has been demonstrated to converge to a global optimum, within a finite number of iterations [80]. Due to QC's probabilistic nature, the QC step does not always guarantee a solution with less than 100% optimality gap. It should be noted that the proposed method is heuristic in nature and always converges provided that a feasible solution exists at each step of the hybrid QC-IQFP parametric method. The proposed hybrid QC-IQFP parametric method uses the inexact parametric algorithm [80] as a support framework combined with QC-based solution techniques to solve the vehicle routing problems. Since a global optimal solution is not required at each step of the parametric algorithm, a quantum computer can be exploited to obtain near-optimal solution for the formulated QUBO problem at the corresponding step.

## 7.3. Computational Results

Computational experiments are conducted based on vehicle routing instances of various sizes to test the computational efficiency of the proposed hybrid QC-IQFP parametric method. The vehicle routing problem here is an IQFP problem and can be solved using MINLP solvers Bonmin and BARON. It is important to note that BARON is a global optimization solver, meaning that the optimal solutions obtained by BARON are the global optimal values for the corresponding vehicle routing instances. A 24-hour time limit is enforced on both MINLP solvers for a more appropriate comparison with the hybrid parametric method. The classical computer mentioned in Section 3 is used to solve the IQFP instances. The same machine is used as a classical backend for the hybrid QC-IQFP parametric method to perform simple computations and value comparisons. Problems in the QC step are solved on the quantum processor. Additionally, large Hamiltonians corresponding to the QC step are solved using *qbsolv* tool with the same quantum processor as its backend. The sub-QUBOs are compiled and run on the quantum processor reported in Section 3. The computational times for both classical and hybrid quantum procedures were recorded at each step, and the required total time is reported in Table 5 for each considered instance.

The size of the IQFP vehicle routing problem depends on the number of customer locations and available vehicles. It should also be noted that problem sizes exhibit a quadratic growth with the number of customers to be serviced. The problem sizes increase sequentially by changing the number of customer locations and vehicles in an ordered manner to identify any trends with respect to computation time and solution quality. Problem sizes for all instances are also reported in Table 5. Here, the largest vehicle routing instance consists of 12 customer locations and 4 available vehicles, corresponding to the problem size of 676 binary variables and 60 constraints in the quadratic fractional programming problem. Although the QUBO problem in the QC step changes dynamically with each iteration, the size of this problem remains fixed and equal to the size of the original IQFP problem. In order to illustrate the application of this vehicle routing model, we also consider an instance with 5 customer locations to be serviced using 2 vehicles as shown in Figure 8. Solution to this problem yields two optimal routes utilizing both available vehicles as shown in this figure. These optimal routes minimize the logistic ratio, which is the fractional objective function, while making sure that each route begins and ends at the depot. Also as seen in the illustration, each customer is visited only once as enforced by the defined constraints. In this case, both vehicles have been assigned a route, but under-utilization of vehicles is also possible for some vehicle routing problems.

It can be clearly seen that the hybrid QC-IQFP parametric method obtains high quality solutions within reasonable computation times, by comparing the computational times and objective function values reported in Table 5 for each solution strategy. This method also performs better than the MINLP solvers Bonmin and global optimizer BARON by using less computation time for medium and large-sized instances. Even for small vehicle routing problems, the QUBO problem corresponding to the QC step cannot be directly embedded on the quantum computer. The proposed hybrid QC-IQFP parametric method requires longer computation time to solve these vehicle routing problems as the QUBO problem needs to be partitioned into smaller subproblems during each iteration of the hybrid method. Performance of BARON starts deteriorating beyond problem sizes with 128 variables, and Bonmin exhibits poor performance at relatively larger instances. Deviation from the global optimum is also observed in some cases with both Bonmin solver and the hybrid QC-IQFP parametric method, but this deviation lies within 10% from its global optimum. The hybrid QC-IQFP parametric method obtains a solution within 2 hours in case of medium size instances. Similarly, for larger instances, both nonlinear solvers could not find an optimal solution after 24 hours, while the proposed hybrid QC-IQFP parametric method obtains a near-optimal solution within 10 hours of computation time. Unlike any of the previous case studies, the time required by the hybrid strategy reflects the majority of the computation time used by the quantum processor and the time required to partition large QUBO problems.

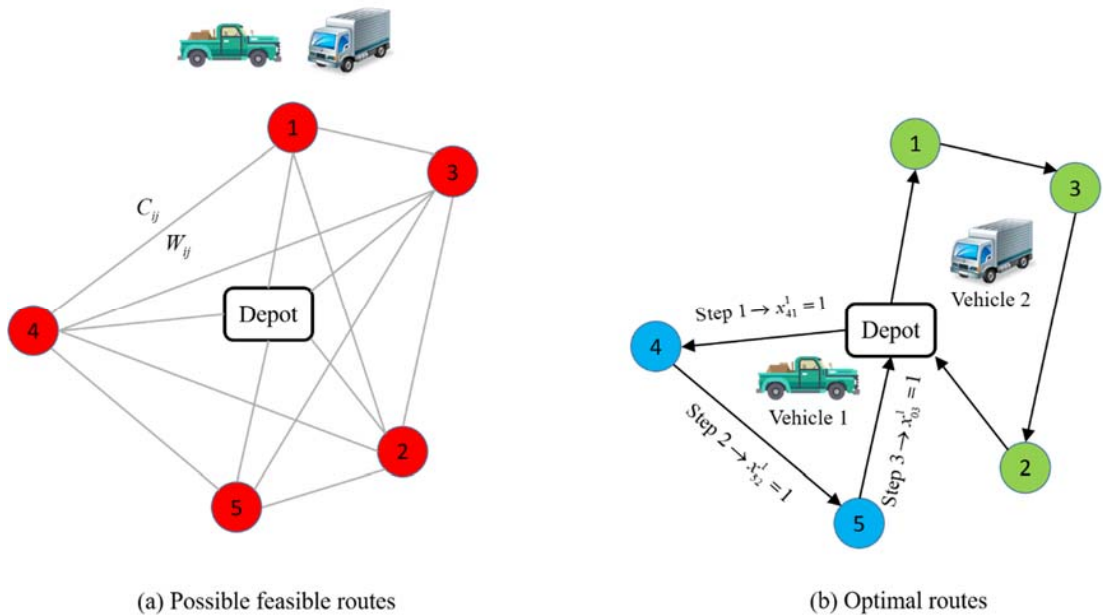

*Figure 8. Vehicle routing problem with two available vehicles showing a) all possible routes and b) optimal routes for each vehicle*

Without any operational constraints like the demand or capacity constraints, this vehicle routing problem is relatively loosely constrained. Despite this fact, classical MINLP solvers face difficulty in tackling quadratic fractional problems. Furthermore, computational performance of the hybrid QC-IQFP parametric method can be improved by using the same embedding scheme for the QUBO problem in the QC step, since the size of the QUBO problem remains fixed throughout the process. Lowering the acceptable tolerance level can also improve the solution quality. QC's probabilistic nature can obstruct the performance of the proposed hybrid QC-IQFP parametric method by impeding the algorithm from reaching convergence. With the growing scalability and improving qubit error-correction schemes, the heuristic nature of the hybrid QC-based method can be subdued to further compete with any tailored solution algorithm for solving IQFP problems. Overall, the hybrid QC-IQFP parametric algorithm performs better than general purpose MINLP solvers in solving large IQFP problems in terms of computational time and solution quality in some cases. This hybrid QC-based method can be used competitively with other MINLP solvers in solving large-scale IQFP problems and would lead to economic benefits when applied to vehicle routing problems.

## 8. Conclusions

In this paper, we proposed hybrid QC-based solution strategies for solving large-scale mixed-integer optimization problems. The applicability of hybrid QC-based algorithms was demonstrated through application problems of practical relevance, namely, the molecular conformation problem, job-shop scheduling problem, manufacturing cell formation problem, and the vehicle routing problem. The molecular conformation problem was reformulated into a QUBO problem and solved directly using the hybrid QC partitioning algorithm. In the second application, the job-shop scheduling problem was solved with a proposed hybrid QC-MILP decomposition method. Moreover, the hybrid QC-MIQP stepwise decomposition method was developed specifically to solve the manufacturing cell formation problem. We further proposed a hybrid QC-IQFP parametric method to solve the vehicle routing problem. The computational results showed that the proposed hybrid QC-based algorithms clearly outperformed general-purpose state-of-the-art exact solvers for solving large-scale mixed-integer optimization problems. Although the exact solvers were efficient in solving small-scale problems, a clear quantum advantage was perceived with hybrid QC-based solution techniques for large-scale optimization problems. The performance of the proposed hybrid QC-based solution strategies was independent of the annealing-based QC device used to perform quantum computations, and may improve with the scalability of such devices.

*Table 5. Computational results of the vehicle routing problems*

| Customer locations | Vehicles | Binary variables | Bonmin 15 | | Baron 19.7.3 | | Hybrid QC-IQFP parametric method | | |
|---|---|---|---|---|---|---|---|---|---|
| | | | Time (s) | Min obj. | Time (s) | Min obj. | Iterations | Time (s) | Min obj. |
| 3 | 2 | 32 | 1.80 | 0.68 | 1.04 | 0.68 | 3 | 0.14 | 0.68 |
| 5 | 2 | 72 | 15 | 0.5 | 7 | 0.5 | 3 | 497 | 0.50 |
| 6 | 2 | 98 | 46 | 0.49 | 44 | 0.45 | 4 | 682 | 0.45 |
| 7 | 2 | 128 | 103 | 0.40 | 1630 | 0.40 | 3 | 1,132 | 0.4 |
| 8 | 3 | 243 | 2,124 | 0.26 | 86,400* | 0.26 | 3 | 924 | 0.26 |
| 9 | 3 | 300 | 7,921 | 0.21 | 86,400* | 0.21 | 4 | 1,059 | 0.21 |
| 10 | 3 | 363 | 15,590 | 0.40 | 86,400* | 0.37 | 5 | 3,415 | 0.40 |
| 11 | 3 | 432 | 36,246 | 0.32 | 86,400* | 0.31 | 15 | 5,813 | 0.33 |
| 11 | 4 | 576 | 86,400* | -- | 86,400* | 0.27 | 4 | 3,573 | 0.28 |
| 12 | 4 | 676 | 86,400* | -- | 86,400* | 0.24 | 77 | 35,299 | 0.28 |

* Timeout of 24 hours (86,400 CPUs) reached and no solution is returned.

## Acknowledgements

This research used resources of the Oak Ridge Leadership Computing Facility, which is a DOE Office of Science User Facility supported under Contract DE-AC05-00OR22725. This work is supported in part by the U.S. Department of Energy, Office of Science, Early Career Research Program.

## Nomenclature

### Molecular Conformations

*Parameters*

| | |
|---|---|
| $\beta$ | Penalty parameter for each bond pair |
| $\varepsilon$ | Depth of Leonard-Jones potential well |
| $l_b$ | Bond length |
| $r_{ij}$ | Distance between locations of atoms $i$ and $j$ |
| $\sigma$ | Leonard-Jones pairwise distance at which potential is minimum |
| $U_{ijkl}$ | Potential energy contribution due to atoms $i$ and $k$ at locations $j$ and $l$, respectively |
| $U_{ij}^{LJ}$ | Potential energy contribution due to pairwise interaction between atoms $i$ and $j$ |
| $U_{ij}^{bond}$ | Potential energy contribution due to presence of bonds between atoms $i$ and $j$ |

*Binary variables*

| | |
|---|---|
| $x_{ij}$ | Binary variable that indicates whether atom $i$ is placed at location $j$ |

### Job-shop Scheduling

*Sets*

| | |
|---|---|
| $I$ | set of jobs or orders |
| $M$ | set of machines |

$C_{im}$                Processing cost of job $i$ on machine $m$

$D_i$                Due date for processing job $i$

$P_{im}$                Processing time for job $i$ on machine $m$

$R_i$                Release date of job $i$

*Binary variables*

$x_{im}$                Binary variables that indicates whether job $i$ is processed on machine $m$

$y_{ij}$                Binary variable denoting whether job $j$ is processed after job $i$ on the same machine

*Continuous variables*

$ts_i$                Start time of job $i$

## Manufacturing Cell Formation

*Sets*

$M$                set of machines

$P$                set of parts

$R$                set of cells

*Parameters*

$a_{ij}$                Whether part $i$ requires machine $j$

$c_i$                Inter-cell movement cost per unit of part $i$

$o_{ij}$                Number of times part $i$ requires operation on machine $j$

$u_{ij}$                Cost of part $i$ not utilizing machine $j$

$v_i$                Number of units of part $i$

*Binary variables*

$y_{jk}$                Binary variable that indicates whether machine $j$ is in cell $k$

*Continuous variables*

$x_{ik}$                Whether part $i$ is processed on cell $k$

# Vehicle Routing